# A Novel Agent Based Approach for Controlling Network Storms

[1]Dr. T.R.Gopalakrishnan Nair (SMIEEE), [2]B.R.Shubhamangala, [3]Vaidehi.M (MIEEE)

[1]Director, Research and Industry Incubation Centre, Dayanandasagar Institutions,
Bangalore, India
trgnair@ieee.org

[2]Associate member, Research and Industry Incubation Centre, Dayanandasagar Institutions
Bangalore, India
brm1shubha@gmail.com

[3] Research Associate, Research and Industry Incubation Centre, Dayanandasagar Institutions
Bangalore, India
dm.vaidehi@gmail.com

*Abstract*-One of the fundamental data transmission mechanisms in Ethernet LAN is broadcasting. Flooding is a direct broadcasting technique used in these networks. A significant drawback of this method is that it can lead to broadcast storms. This phenomenon is more common in multivendor switch environment. Broadcast storms usually results in dissension, collision and redundancy leading to degradation of the network performance. Most of the storms appear without much warning and it affects the efficiency of network even in situations when the network is expected to work most efficiently. There are several characteristic patterns by which storm can appear in a LAN, like rate monotonic repetition, transient appearances with different types of growth properties and decay profiles.
In this paper we discuss the storm build up pattern in an industry and present various reasons for storm in LAN. We have identified a strategy for controlling network storms, using multiple static agents. These agents inhibit storm packet regeneration in the network using the knowledge of storm growth pattern. A model developed out of empirical studies is used to differentiate normal packet growth from storm packet growth and used in control mechanism of storms

## I. Introduction

Broadcast is one of the technique for multipoint delivery of messages in Ethernet LAN [1][2]. In Ethernet LAN some protocols such as ARP, OSPF, NLSP, NETBIOS, NETBEUI, SMB, DHCP, RIP1, SAP and IPX work in broadcast mode for effective network operation, such as reporting the status of network and event signaling. When LAN is under broadcast mode, broadcast traffic is of vital concern. When broadcast traffic is less, it does not harm the performance of the network. The normal rule is that broadcast traffic must not exceed 20% of the total traffic for safe network operation. Extensive broadcast traffic in the network, leads to a serious adverse situation called broadcast storm. Broadcast storms are usually caused by protocol events such as ARP request or by other problems in the network such as faulty interface, or network loops. When a device desires to locate a target server, it sends a sequence of ARP packets. If the sever device fails to respond or sender device is having a problem in broadcast mechanism, the source device will not get any response for the broadcast. At that time, the source device starts endlessly broadcasting the packets leading to loop ARP sequence [3]. Such an event eventually leads to broadcast storm in the network. Some of the reasons for broadcast storm in medium level enterprise are faulty cabling, misconfigured protocols like NetBIOS/NetBEUI, incorrect network design and plan, faulty NIC cards, loop in the network, and loops in multivendor switch environment and virus.

Though network failure is causing severe problems in enterprises and various causes for network failure due to storm are well known, no effective mechanism has been positioned to control the storm in medium level enterprise LAN. Here we attempt to characterize and improve storm control situation by employing multiple static agents in each node of the network.

The nature of growth of normal broadcast traffic is fed to multiple static agents [15]. When the network is active, static agents monitor the sequence of growth of broadcast packets. If the nature of current traffic buildup is different from normal growth and is reaching the threshold value, it triggers the control mechanism which in turn controls the growth of broadcast packets. The organization of the paper is as follows, in section II the background on the various reasons of broadcast storm in the LAN[16]-[18] is presented. Section III is about storm pattern analysis. Section IV projects the strategy to control the storm. Finally section V presents the conclusion and future work.

## II Background

Ethernet LAN consists of a set of nodes that may communicate with one and another from time to time. In this environment a node may communicate with other nodes in the network directly sometimes through flooding. In a typical situation it can lead to occurrence of broadcast storms. The reasons of broadcast storms to occur are mentioned below.

*A. Reasons for broadcast storm*

Faulty NIC

A faulty NIC may transmit a large amount of junk data packets endlessly causing broadcast storm in the network. For example if faulty card transmits ARP at a very high rate of several hundred frames per second, request can be propagated to other devices eventually leading the network towards storm [2]. A faulty NIC may also jabber the packets into network endlessly. This jabbering creates false status of network busy and there by stopping all other traffic on the network. Endless jabbering also leads to network storms.

Virus

Virus such as smurf and fraggle, work by sending spoofed packets such as ICMP echo replies, UDP echo replies and UDP echo packets to broadcast addresses, making all the nodes reply to the spoofed address which floods the network heavily. This in turn causes broadcast storm.

Loops

Loops may get created in the network accidentally [4]. Loops leads to severe broadcast storms. Broadcast storm builds up very fast in loops compared to other causes. If STP is enabled on switch, loops can be successfully avoided.

Multivendor Switch Environment

In the medium level enterprise LAN, it is common to have switches from multivendors. In such situation if loop gets created because of redundant link or because of physical cable connection, though STP is enabled on all switches, vulnerability of network towards broadcast storm is 80%. This practical problem in the enterprise. Due to multivendor incompatibility or ambiguous protocol specification, loop detection and avoiding mechanism may not work. This also leads to the network towards storm.

*B. Stages of Broadcast Storm*

Broadcast storm phenomena can be divided into three stages namely

1. Initial stage

In this stage the network operations are timed out and slow response of network is observed. Bandwidth consumption leads to network time outs and slow response.

2. Storm buildup stage

This is a practical problem in the enterprise. Here users cannot access the server and intermittent operation of the network is observed.

3. The final and advanced stage

In this stage storm results in total network meltdown due to severe congestion and packet collision. Here 90% of the bandwidth is consumed by the storm [5]. At this point, the transmission of broadcast packets develops into an interesting pattern where packets build up slowly in the initial stage and raise very fast within few milliseconds; it reaches the peak and suddenly drops down. This pattern repeats. The graph of practical wave forms obtained is shown in the below figure1. Sample data of obtained waveform is given in Table 1.

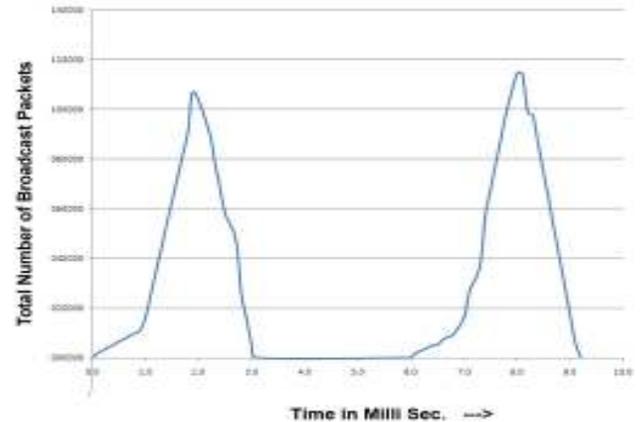

Figure 1A Graph of practically obtained broadcast storm.

Table1 Sample data of practically obtained wave form

| Time in ms | No of broadcast packets | Time in ms | No of broadcast packets |
|---|---|---|---|
| 0 | 0 | 1.5 | 63300 |
| 0.1 | 1900 | 1.6 | 72700 |
| 0.2 | 3200 | 1.7 | 82000 |
| 0.3 | 4200 | 1.8 | 91300 |
| 0.4 | 5200 | 1.9 | 107200 |
| 0.5 | 6600 | 2.2 | 90720 |
| 0.6 | 7600 | 2.3 | 79000 |
| 0.7 | 8900 | 2.4 | 68400 |
| 0.8 | 9900 | 2.5 | 57900 |
| 0.9 | 11100 | 2.7 | 47600 |
| 1 | 16400 | 2.8 | 26300 |
| 1.1 | 25900 | 2.9 | 15800 |
| 1.2 | 35300 | 3 | 5200 |
| 1.3 | 44500 | 3.1 | 0 |
| 1.4 | 54300 | | |

The list of variables used in this paper are given in Table 2

Table 2 List of variables

| List of Variables | Notation Description |
|---|---|
| P | Size of packet of which node is transmitting. |
| I | The transmission interval, which is a constant. |
| $t_{start}$ | Starting time of transmission |
| $t_{end}$ | End time of transmission |
| Ps | PTR measured at $t_{start}$ |
| Pe | Safe threshold value of PTR measured at $t_{end}$ |
| m | Sampling rate |
| t | Instantaneous time |
| P(t) | Instantaneous PTR at the time t. |

*C. Metrics of Broadcast Storm*

Important parameters of network broadcast storm are mentioned here. These parameters are used to control the broadcast storms.

1. Bandwidth

The maximum number of packets that can be transmitted in the Ethernet is calculated. If current number of broadcast packets transmitted is near to maximum number of packets that can be transmitted, it indicates high bandwidth consumption due to storm[5][6]. For normal broadcast operation, the broadcast traffic must be within 20% compared to the total traffic. Apart from this bandwidth consumption of each node is denoted by $N_{BW}$ and is estimated as follows.

$N_{BW} = P*I$ .......( 1)

Where P = The size of the packet of which the node is transmitting.

I = The transmission interval, which is a constant.

This constant value depends on the IPG (Internet Packet Gap) value of the channel. If current $N_{BW}$ value is much greater than the normal permissible value of the $N_{BW}$, then it indicates that, this particular node is storming the network.

2. Inter Packet Gap

For collision free transmission of frames, Ethernet allocates minimum idle time gap at the end of each frame. This gap is called as inter packet gap (IPG) or more appropriately as inter frame gap (IFG) [7]. This idle time is utilized by network media for its stabilization and other network devices utilize this time to process the frame. This solely depends on the capacity of Ethernet. The minimum IPG is 96 bits /seconds which is 9.6micro seconds for 10 Mb/s Ethernet, 960 nano seconds for 100 Mb/s Ethernet, 96 nano seconds for 1Gb/s Ethernet and 9.6 nano seconds for 10 Gb/s Ethernet. If collision is occurring in the network, deep shrinkage in the IPG is observed [7]. This is an important parameter to detect collision in the channel. High collision indicates broadcast storm.

3. Utilization

Network utilization is the ratio of current network traffic to maximum traffic the Ethernet can handle. Utilization is the indication of bandwidth consumption in the network. High utilization percentage indicates network busy status where as utilization below10% indicates idle status of the network [8]. High utilization percentage is mainly dependent upon network type, topology and other network parameters. For a switched Ethernet if utilization is 50%, it is considered as having a good efficiency. If utilization percentage is more than 60%, conclusion is network is suffering from broadcast storm.

4. IPID

Uniquely IPID indicates the packet flow. IPID can be observed in the network monitoring tools [9], [10]. If same IPID is repeated, it indicates storm in the network due to loops. This aspect is very important while working in multivendor switch environment.

III. STORM PATTERN ANALYSIS

To develop a mechanism to control storms, mathematical analysis of storm build up is very essential. Only the growth pattern of the storm upto threshold value is important, for our study. Here we analyze the packet transmission rate (PTR) from initial time of transmission (of the sample given) denoted by $t_{start}$ to end time of transmission (of the sample given) denoted by $t_{end}$. Storm raise pattern is given in figure 2. Values are given in table 3.

Let the initial PTR be Ps measured at $t_{start}$ and safe threshold value of PTR be Pe measured at $t_{end}$.

Let p(t) be instantaneous PTR at the time t.

So, P(t) can be expressed as, p

$P(t) = at + bt*e^{mt}$ .........(2)

Where,

m=Sampling rate

t= instantaneous time

$$a = 2\pi * \frac{(Pe - Ps)}{m}$$

$$b = 2\pi * Ps$$

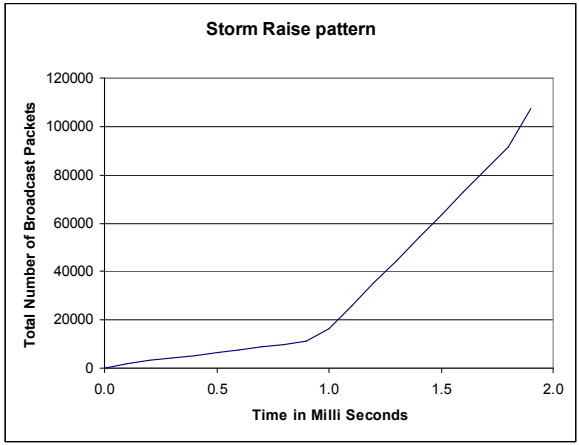

Figure 2. Storm raise pattern

Table 3 Storm raise pattern values

| Time in ms | TNBP | Time in ms | TNBP |
|---|---|---|---|
| 0 | 0 | 1.1 | 25900 |
| 0.1 | 1900 | 1.2 | 35300 |
| 0.2 | 3200 | 1.3 | 44500 |
| 0.3 | 4200 | 1.4 | 54300 |
| 0.4 | 5200 | 1.5 | 63300 |
| 0.5 | 6600 | 1.6 | 72700 |
| 0.6 | 7600 | 1.7 | 82000 |
| 0.7 | 8900 | 1.8 | 91300 |
| 0.8 | 9900 | 1.9 | 107200 |
| 0.9 | 11100 | 2.2 | 90720 |
| 1 | 16400 | 2.3 | 79000 |

Where TNBP= Total number of broadcast packets

## IV STRATEGY TO CONTROL STORM

Equation 2 is used to calculate instantaneous Packet Transmission Rate (PTR) during the normal broadcast phenomena from pstart to $p_{end}$ at an interval of one millisecond. This process gives an array of normal broadcast PTR. It can be referred as PTR array. If PTR array points are plotted, it gives a normal broadcast wave pattern from $P_{start}$ to $P_{end}$.

In order to control the storm, we are propose an agent based system that is static multi agent system. Our proposed system is on compliant with IEEE FIPA system both at FIPA agent level and platform level [11][19]. Multi agent system consists of multiple distributed autonomic static agents. This feature gives high degree of flexibility, adaptability towards larger network and greater ability in a dynamic environment [12] [13]. The proposed agents have the following characteristics. Agent is a software module that can be embedded in each node of network. Agent executes the required operation independently. Agents can use the resource that is available in node and also in the network [14] [20].

Agent software collects the operational data (e.g., statistics such as number of broadcast packets, bandwidth consumption) and detects exceptional events such as number broadcast packets reaching threshold value and takes appropriate action based on its knowledge base. These multi agents cater to various challenges in storm control and network management such as quick analysis, handling unanticipated changes in the network, controlling storm packet transmission while maintaining normal operation of the network.

The Autonomic static agents are employed at each node of the network as shown in the figure 3. Our proposed model of agent system is subdivided into three models namely Communication agent model, Storm handler agent model and Control agent model as shown in the figure 4.

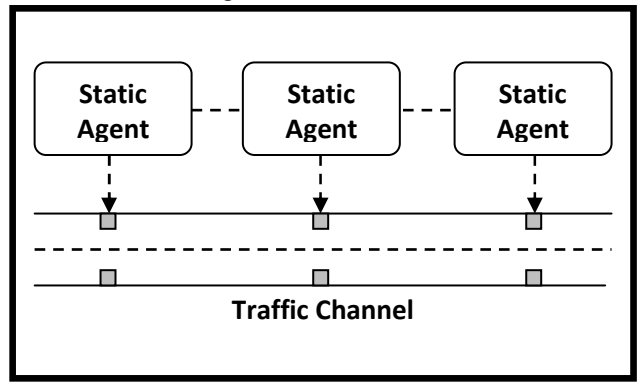

Figure 3 Static agent at each node of network

Figure 4 Architecture of static node

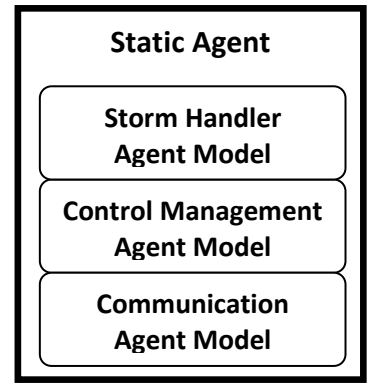

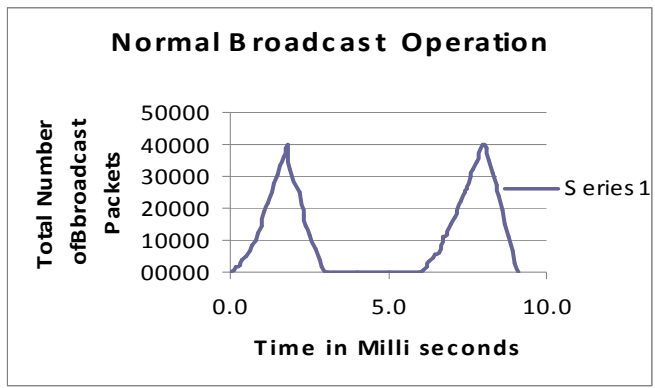

Figure 5 Normal broadcast operation

Table 4 Normal broadcast operation sample values

| Tms | TNBP | Tms | TNBP | Tms | TNBP | Tms |
|-----|------|-----|------|-----|------|-----|
| 0   | 0    | 1.5 | 30800| 5   | 0    | 7.6 |
| 0.1 | 400  | 1.6 | 33400| 6   | 0    | 7.7 |
| 0.2 | 1400 | 1.7 | 36000| 6.1 | 800  | 7.8 |
| 0.3 | 2400 | 1.8 | 40000| 6.2 | 1800 | 8   |
| 0.4 | 3750 | 1.9 | 32000| 6.3 | 3100 | 8.1 |
| 0.5 | 5000 | 2.2 | 25000| 6.4 | 4200 | 8.2 |
| 0.6 | 6400 | 2.3 | 18000| 6.5 | 5700 | 8.3 |
| 0.7 | 8150 | 2.4 | 14400| 6.6 | 6200 | 8.4 |
| 0.8 | 10000| 2.5 | 11800| 6.7 | 10800| 8.5 |
| 0.9 | 12500| 2.7 | 7000 | 6.8 | 11000| 8.6 |
| 1   | 14650| 2.8 | 4000 | 7   | 15000| 8.7 |
| 1.1 | 18760| 2.9 | 1800 | 7.1 | 17000| 8.8 |
| 1.2 | 21800| 3   | 0    | 7.3 | 22000| 8.9 |
| 1.3 | 24800| 3.1 | 0    | 7.4 | 25000| 9   |
| 1.4 | 28000| 4   | 0    | 7.5 | 26200| 9.1 |

*Functioning of each mode of the architecture of figure 4 is described below.*

A. Communication agent model

This agent has direct contact with the Ethernet Channel. It reads the real-time traffic data in the network for every millisecond. Once data is collected, it transports data to Control agent model.

B. Control agent model

Initially during the normal broadcast operation, control management agent requests communication agent model to get network statistics. Once data is received, control management agent calculates the growth pattern of packets up to threshold value using equation 2 and is stored in the array called PTR array. Apart from this, Control management agent has the data base of threshold values of other vital parameters of network such as Bandwidth, IPG and Utilization. Once the data base is ready, it sends signal to communication model, to start capturing the data for every millisecond. Communication agent model starts working only after receiving this signal.

When control model starts getting the data, it calculates instantaneous PTR using the equation 2 and stores in the array called PTR compare array. PTR compare array is compared with PTR array which has normal broadcast pattern points. If deviation is less than 5%, no action is taken and network operation will be normal. If PTR compare array deviates more than 5%, it intimates the storm handler agent model that broadcast storm is going to occur and it also identifies the node where the storm is originated.

C. Storm handler agent model

When trigger signal is obtained from control agent model, particular node gets blocked and will not allow any broadcast packets in the network through that node and will generate a trouble ticket to notify broadcast storm in that particular node. This gives an opportunity for network administrator to know the occurrence of broadcast storm immediately. Once the fault has been removed, node can be reconnected to network. Thus broadcast storms can be controlled by multiple static agents. Algorithm 1 presents the procedure how static agent control the storm.

Algorithm1:

Input: channel characteristics such as number of broadcast packets, bandwidth consumed, utilization
Output: Number of broadcast packets is restricted below the threshold.
Communication agent model: read the channel characteristics under normal operation. Transfer to Agent control model
Agent control model: Calculate safe range of PTR array, IPG, bandwidth, number of broadcast packets.
Trigger to communication agent model to start capture the channel characteristics.
 If trigger ==1
Start capturing the statistics.
 Transfer to control agent model.
Control agent model: If statistics >threshold value,
Generate trap==error
Send to Storm handler model(SHM)

SHM: If trap== error ==1,
Stop broadcast packets in that node.
Generate the trouble ticket to notify the error.
 Continue till the error is rectified.

Simulation results of storm control model graph are given in figure 6 and corresponding sample values are given in the Table 5.

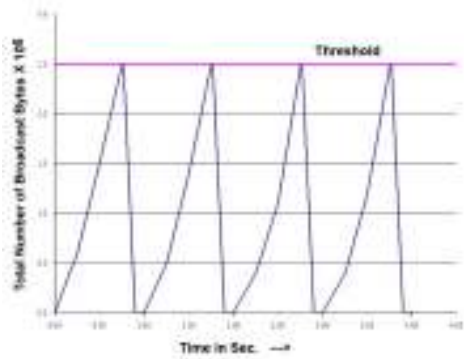

Figure 6 Storm control model graph

Table 5 Storm control simulation values

| T in Secs | TNBP | Threshold | T in Secs | TNBP | Threshold |
|---|---|---|---|---|---|
| 0.00 | 0.0 | 2.5 | 2.10 | 0.0 | 2.5 |
| 0.10 | 0.0 | 2.5 | 2.25 | 0.4 | 2.5 |
| 0.25 | 0.6 | 2.5 | 2.50 | 1.1 | 2.5 |
| 0.50 | 1.5 | 2.5 | 2.75 | 2.5 | 2.5 |
| 0.75 | 2.5 | 2.5 | 2.78 | 2.5 | 2.5 |
| 0.77 | 2.5 | 2.5 | 2.90 | 0.0 | 2.5 |
| 0.90 | 0.0 | 2.5 | 3.00 | 0.0 | 2.5 |
| 1.00 | 0.0 | 2.5 | 3.10 | 0.0 | 2.5 |
| 1.10 | 0.0 | 2.5 | 3.25 | 0.4 | 2.5 |
| 1.25 | 0.5 | 2.5 | 3.50 | 1.2 | 2.5 |
| 1.50 | 1.4 | 2.5 | 3.70 | 2.5 | 2.5 |
| 1.75 | 2.5 | 2.5 | 3.74 | 2.5 | 2.5 |
| 1.78 | 2.5 | 2.5 | 3.90 | 0.0 | 2.5 |
| 1.90 | 0.0 | 2.5 | 4.00 | 0.0 | 2.5 |
| 2.00 | 0.0 | 2.5 | -- | -- | -- |

TNBP = Total number of Broadcast data in MB

The suppression of Broadcast storm can be achieved through hardware or software method.

A. Hardware suppression : Bandwidth based approach

In this approach if the number of incoming broadcast packets on a port within a one second time interval exceeds the configured threshold, the switch filters out all incoming broadcast packets on the port for the remainder of the one second period. This approach measures broadcast activity relative to the total available bandwidth. In our paper we have used the software approach to suppress the storm.

B. Software Suppression : Packet-based Broadcast approach

In this approach if the number of incoming broadcast within one second time interval exceeds the configured threshold, the agent filters out all incoming traffic on the port for the remainder of one second period. Since the packet size varies in general we take the broadcast data in terms of megabytes (MB)

Figures 7 and 8 represents the snapshots of the broadcast storm and normal broadcast operation respectively.

V. CONCLUSION AND FUTURE WORK

This paper reports our research on A novel Static Agent based Approach for Controlling Network Storms. Here, we have analyzed various causes of broadcast storm in a LAN. Storm build up pattern is practically observed and based on this observation an equation for storm build up is formulated. An IEEE FIPA compliant multiple static agent based system is identified to suppress the network storm. Agent software collects the operational data (e.g., statistics such as number of broadcast packets, bandwidth consumption) and detects exceptional events such as number of broadcast packets reaching threshold value. It takes appropriate suppression action based on its knowledge base, while maintaining normal operation of network. An alert is given about the storm to occurring port. The proposed solution improves the performance of network. Future work can involve static agent response studies and various functional variation of threshold judgment.

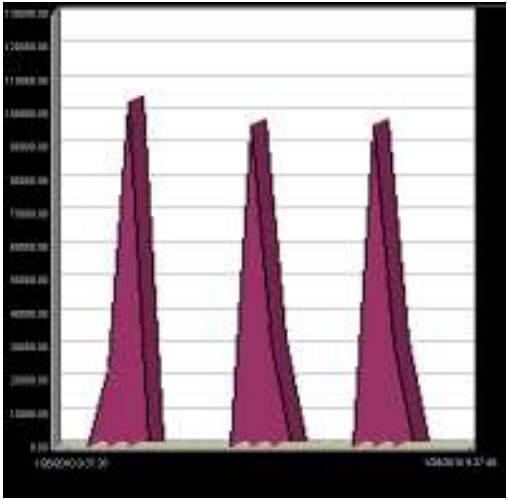

Figure 7. Broadcast storm snapshot obtained from Etherpeak (Network monitoring tool)

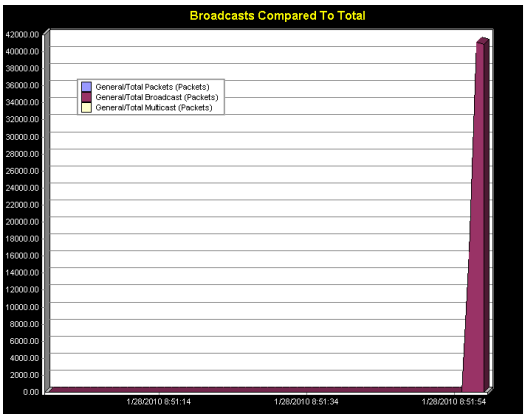

Figure 8 Normal Broadcast operation snapshot obtained from Etherpeak (Network monitoring tool)

## 4. ACKNOWLEDGEMENT

Our acknowledgement is due to Mr. Jagadish Aradhya Senior Manager, Network- IT, Aegis Tech Ltd., Bangalore, for his support throughout. We thank all other experts who provided invaluable suggestions in the conceptualization of the project.